\documentclass[aps,prl,twocolumn,superscriptaddress,showpacs,letter]{revtex4}

\usepackage{graphicx}

\newcommand{\undoped}{BaFe$_2$As$_2$}
\newcommand{\Co}{Ba(Co$_{0.06}$Fe$_{0.94}$)$_2$As$_2$}
\newcommand{\K}{Ba$_{0.6}$K$_{0.4}$Fe$_2$As$_2$}
\newcommand{\Ef}{$E_F$}

\begin{document}


\title{Electronic Structure of the \undoped~Family of Iron Pnictides}


\author{M. Yi}
\author{D. H. Lu}
\affiliation{Stanford Institute for Materials and Energy Sciences, SLAC National Accelerator Laboratory, 2575 Sand Hill Road, Menlo Park, CA 94025, USA}
\affiliation{Geballe Laboratory for Advanced Materials, Departments of Physics and Applied Physics, Stanford University, CA 94305, USA}
\author{J. G. Analytis}
\author{J.-H. Chu}
\affiliation{Geballe Laboratory for Advanced Materials and Department of Applied Physics, Stanford University, CA 94305, USA}
\author{S.-K. Mo}
\affiliation{Geballe Laboratory for Advanced Materials, Departments of Physics and Applied Physics, Stanford University, CA 94305, USA}
\affiliation{Advanced Light Source, Lawrence Berkeley National Laboratory, Berkeley, CA 94720, USA}
\author{R.-H. He}
\affiliation{Stanford Institute for Materials and Energy Sciences, SLAC National Accelerator Laboratory, 2575 Sand Hill Road, Menlo Park, CA 94025, USA}
\affiliation{Geballe Laboratory for Advanced Materials, Departments of Physics and Applied Physics, Stanford University, CA 94305, USA}
\author{X. J. Zhou}
\author{G. F. Chen}
\author{J. L. Luo}
\author{N. L. Wang}
\affiliation{Beijing National Laboratory of Condensed Matter Physics, Institute of Physics, Chinese Academy of Sciences, Beijing 100080, China}
\author{Z. Hussain}
\affiliation{Advanced Light Source, Lawrence Berkeley National Laboratory, Berkeley, CA 94720, USA}
\author{D. J. Singh}
\affiliation{Materials Science and Technology Division, Oak Ridge National Laboratory, Oak Ridge, TN 37831-6114, USA}
\author{I. R. Fisher}
\affiliation{Geballe Laboratory for Advanced Materials and Department of Applied Physics, Stanford University, CA 94305, USA}
\author{Z.-X. Shen}
\email{zxshen@stanford.edu}
\affiliation{Stanford Institute for Materials and Energy Sciences, SLAC National Accelerator Laboratory, 2575 Sand Hill Road, Menlo Park, CA 94025, USA}
\affiliation{Geballe Laboratory for Advanced Materials, Departments of Physics and Applied Physics, Stanford University, CA 94305, USA}

\date{\today}

\begin{abstract}
We use high resolution angle-resolved photoemission spectroscopy to study the band structure and Fermi surface topology of the \undoped~iron pnictides. We observe two electron bands and two hole bands near the X-point, $(\pi,\pi)$ of the Brillouin zone, in the paramagnetic state for different doping levels, including electron-doped \Co, undoped \undoped, and hole-doped \K. Among these four bands, only the electron bands cross the Fermi level, forming two electron pockets around X, while the hole bands approach but never reach the Fermi level. We show that the band structure of the \undoped~family matches reasonably well with the prediction of LDA calculations after a momentum-dependent shift and renormalization. Our finding resolves a number of inconsistencies regarding the electronic structure of pnictides.
\end{abstract}

\pacs{74.25.Jb, 74.70.-b, 79.60.-i}

\maketitle


One reason the Fe-pnictides have captured and retained so much attention in the high T$_c$ field is the complexity of its electronic structure and the many mysteries that still remain unsolved. The undoped parent compound, \undoped, exhibits a collinear antiferromagnetic spin-density-wave (SDW) ordering~\cite{rotter,delacruz,huang}, which is suppressed with doping leading to superconductivity~\cite{rotterdoped,chen,ni,chu}. It has been suggested based on LDA calculations that the SDW results from nesting of the hole-like Fermi surface (FS) at $\Gamma$ and electron-like FS at X connected by the wavevector ($\pi, \pi$)~\cite{fang,singh}. However, angle-resolved photoemission spectroscopy (ARPES) reports so far do not seem to converge on the electronic structure at the X-point~\cite{feng,kaminski,zhouspots,zhouPG,ding,borisenko,dingK,hasan,dingCo}, and also appear to be inconsistent with LDA calculations, painting a different picture from the earlier work on other pnictide compounds~\cite{lu}. Given the complexity of the multiband nature of the material, it is important to have a good understanding of the basic character of the band structure before more subtle many-body physics can be understood.

In this paper, we present a detailed high resolution ARPES study to clarify the electronic structure around the X-point in the family of \undoped~compounds in the paramagnetic state. At temperatures above all transitions in these materials, we resolve two electron-like dispersions that cross the Fermi level (\Ef)~and two hole-like dispersions that approach but remain below \Ef. Furthermore, from a series of parallel cuts across the region at the Brillouin zone (BZ) boundary near the X-point, we observe that the electron and hole-like bands hybridize with a sizable gap in the region away from the $\Gamma$-X high symmetry line, the details of which are well captured by nonmagnetic LDA calculations after a momentum-dependent shift and bandwidth renormalization, in contrast to the interpretation of a recent ARPES report~\cite{borisenko}. Finally, with measurements on \undoped, \K, and \Co, we show that such a band structure character at X is general in the \undoped~family across both electron and hole doping regimes. Our findings resolve a number of inconsistencies, and provide a coherent picture on the electronic structure of Fe-pnicides.

\begin{figure}
\includegraphics[width=0.47\textwidth]{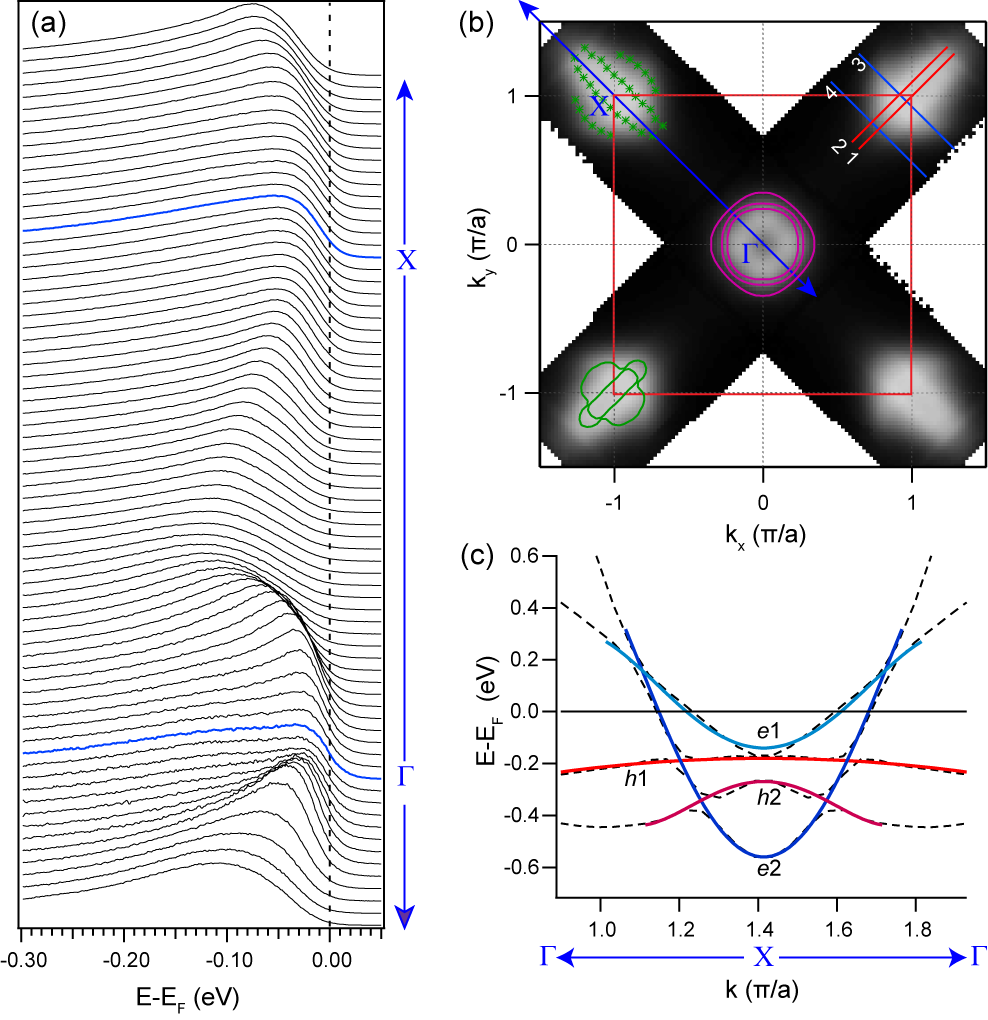}
\caption{\label{fig:fig1}ARPES measurements on \undoped~taken at 140 K ($h\nu$ = 40 eV). (a) EDCs along the high symmetry line $\Gamma$-X marked in (b). (b) FS map taken with an integration window of 5 meV about \Ef, symmetrized to reflect the crystal structure symmetry. The red square marks the 1$^{st}$ BZ. Top left quadrant shows in green the outline of the electron pockets obtained from $k_F$ points. Green lines in the lower left quadrant and magenta lines in the center mark the LDA calculated electron and hole pockets at X and $\Gamma$ respectively, after a nonuniform shift of the \Ef~(down by 0.14 eV at X, and up by 0.01 eV at $\Gamma$). (c) LDA bands near X along the $\Gamma$-X-$\Gamma$ direction shown in dashed lines, resulting from hybridization of two hole band ($h_1$, $h_2$) and two electron bands ($e_1$, $e_2$).}
\end{figure}

High quality single crystals of \K~(T$_c$ = 37 K)~\cite{Kdoped}, \undoped, and \Co~(T$_c$ = 25 K)~\cite{chu} were all grown using the self-flux method. ARPES measurements were taken at both ALS beamline 10.0.1 and SSRL beamline 5-4, with an energy resolution better than 16 meV, and angular resolution of 0.3$^{\circ}$. All samples were cleaved in situ at various temperatures discussed below, and measured under a base pressure better than 4x10$^{-11}$ torr.

The parent compound \undoped~undergoes both an SDW and structural transition at 137 K. The electronic structure in the low temperature state is more complex due to the presence of the SDW order, which will be discussed in details in another paper. In this paper, we focus on the electronic structure in the paramagnetic state above this transition temperature in comparison with nonmagnetic LDA. A FS intensity map of \undoped~taken at 140 K is shown in Fig.~\ref{fig:fig1}(b). Fig.~\ref{fig:fig1}(a) shows Energy Distribution Curves (EDCs) along the high symmetry cut $\Gamma$-X. At $\Gamma$, there is a clear hole-like dispersion which forms a circular hole pocket on the FS, in good agreement with LDA calculations~\cite{singh}. At X, surprisingly, there also appears to be a hole-like dispersion and a cross-shaped FS which may seem to be in contradiction with the two electron pockets predicted by LDA, as also reported by previous ARPES measurements~\cite{feng,kaminski,zhouspots,zhouPG,ding,borisenko,dingK,hasan,dingCo}. However, as a more systematic comparison between the calculated band structure and experimental data in the region near X would show, LDA does not fail as miserably as it has been proclaimed~\cite{borisenko}.

\begin{figure}
\includegraphics[width=0.48\textwidth]{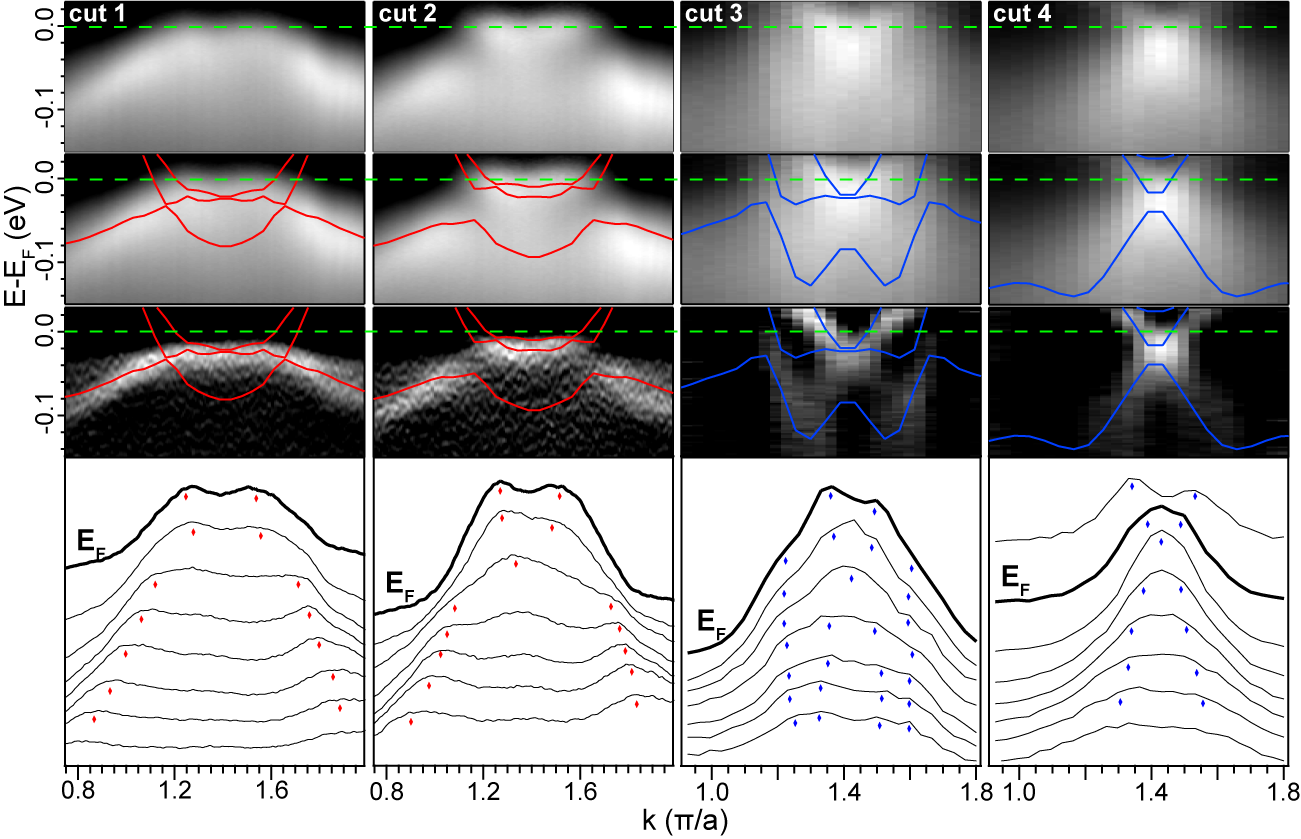}
\caption{\label{fig:fig2} ARPES spectra of \undoped~taken at 140 K ($h\nu$ = 40 eV) along four cuts near X, with cut directions labeled in Fig.~\ref{fig:fig1}(b). The light polarization is perpendicular to to the $\Gamma$-X direction in all four cuts. (a) ARPES spectra divided by the Fermi-Dirac function to reveal band dispersions above \Ef. (b) LDA bands ($k_z$ = $\pi$)~\cite{kz} overlaid on corresponding cuts in (a) after a shift of the \Ef~down by 0.14 eV and a bandwidth renormalization of a factor of 1.5. (c) Second derivative plots of the raw intensity spectra to pull out weak dispersions, overlaid with LDA bands. (d) Momentum Distribution Curves (MDCs) of the raw spectra for the corresponding cuts, marking band dispersions.}
\end{figure}

Let us first take a closer look into the LDA band structure calculated for the paramagnetic state in the vicinity of the X-point. In addition to two electron bands ($e_1$, $e_2$), there are also two hole bands ($h_1$, $h_2$) at higher binding energies (Fig.~\ref{fig:fig1}(c)). The lower electron band $e_2$ intersects both hole bands, and the upper electron band $e_1$ kisses the upper hole band $h_1$ at X. The hybridization between $h_1$ and $e_2$ is forbidden along the exact high symmetry line $\Gamma$-X due to symmetry. However, as we move away from the high symmetry line, the hybridization between these two bands becomes stronger with a sizable gap opening up in the band dispersions. The results imply strong angle dependent variation of the orbital character between xz/yz and xy on the electron FS at X.

\begin{figure*}[t]
\includegraphics[width=0.98\textwidth]{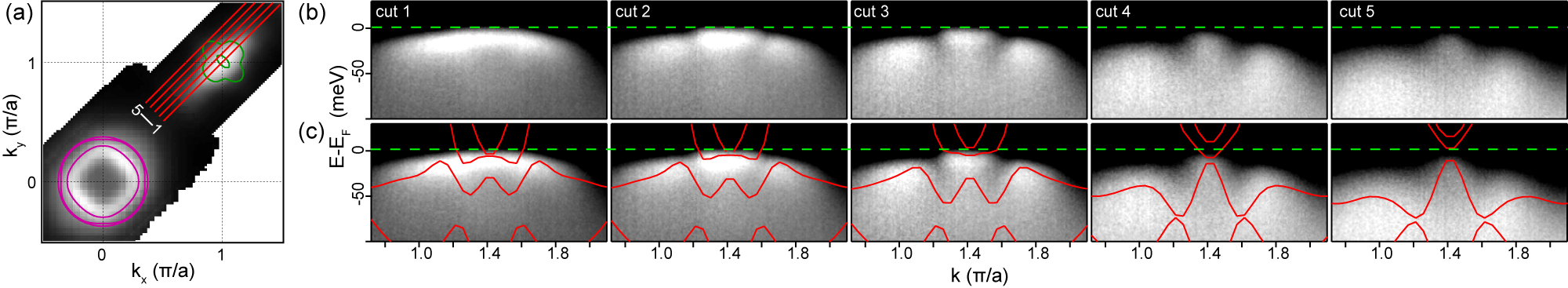}
\caption{\label{fig:fig3}ARPES spectra of \K~(T$_c$ = 37 K) taken at 45 K ($h\nu$ = 27 eV). (a) FS maps around $\Gamma$ and X, overlaid with LDA FS ($k_z$ = 0)~\cite{kz} calculated for \K~after a shift of the \Ef~down by 0.09 eV at X and no shift at $\Gamma$. (b) ARPES spectra along five parallel cuts marked in (a). (c) LDA bands ($k_z$ = 0) for the corresponding cuts, with \Ef~shifted down by 0.09 eV and bandwidth renormalized by a factor of 2.7.}
\end{figure*}

Shown in Fig.~\ref{fig:fig2} are selected cuts in the vicinity of the X-point, two cuts parallel to the $\Gamma$-X direction, and two cuts perpendicular to it. For a better comparison, we superimpose LDA calculations on top of ARPES spectra after a downward \Ef~shift of 0.14 eV and a bandwidth renormalization of a factor of 1.5 (Fig.~\ref{fig:fig2}(b)-(c)). Along the $\Gamma$-X high symmetry line (cut 1), ARPES spectra are dominated by a flat hole-like dispersion. As we go off the high symmetry line (cut 2), we see a clear breakage of this hole band into three segments with the middle segment higher than the sides. This effect grows stronger as we march away from the high symmetry line (not all shown), and is exactly predicted by LDA calculations due to the hybridization between the lower electron band $e_2$ and upper hole band $h_1$. Along the high symmetry cut perpendicular to $\Gamma$-X (cut 3), we see evidence for two electron bands that cross \Ef, as well as the top of a hole-like band in-between, corresponding to the $e_1$, $e_2$, and $h_2$ bands in LDA. The weak lower electron band and hole-like band are more evident in the second derivative plot (row c) and the MDC plot (row d). As we move away from the high symmetry cut (cut 4), the upper electron band gradually disappears above \Ef~while the lower electron band and the hole like band shift upward, as expected from the LDA calculations. Overall, a good agreement can be found between calculated bands and experimental data, including the character (electron or hole) and relative locations of the bands.

One of the reasons that LDA calculation was dismissed as a plausible interpretation in previous ARPES studies~\cite{borisenko} was the failure to observe two electron bands near the X-point. The reason that $e_2$ is only visible in the cut perpendicular to the $\Gamma$-X high symmetry is likely due to a combination of its orbital symmetries, light polarization, and $k_z$ dispersion. Note that all cuts shown in Fig.~\ref{fig:fig2} were taken with an in-plane light polarization perpendicular to the $\Gamma$-X direction, which defines a mirror plane (xz plane in the single-Fe unit cell). Since the light polarization is odd with respect to the mirror plane, bands with even orbital symmetries with respect to this mirror plane ($d_{xz}$, $d_{x^2-y^2}$, and $d_{z^2}$) would be suppressed along this high symmetry line. Therefore, $e_2$ could have one of these orbital symmetries. We note that the observation of two electron bands and hence two electron pockets around X has also been reported in a sister compound SrFe$_2$As$_2$~\cite{hasan}. Therefore, this appears to be general to all undoped 122 parent compounds and can be understood in the framework of LDA calculations.

The comparison with LDA also helps us to understand the FS topology at X. As shown in the cuts parallel to the $\Gamma$-X direction (cuts 1\&2 in Fig.~\ref{fig:fig2}), the $h_1$ dispersion approaches but never crosses \Ef. This can also be confirmed in the cuts perpendicular to the $\Gamma$-X direction (cuts 3\&4 in Fig.~\ref{fig:fig2}). LDA shows that the $e_1$ band bottom kisses the $h_1$ band top. Fig.~\ref{fig:fig2} shows that the $e_1$ band bottom is clearly below \Ef, suggesting that the $h_1$ band top stays below \Ef. Hence at X, only the two electron bands cross \Ef, forming two electron pockets centered at X. In Fig.~\ref{fig:fig1}(b), the top left quadrant shows the outlines of the electron pockets determined from $k_F$  points (momenta at which bands cross \Ef). The two electron pockets hybridize to form an outer pocket and an inner pocket. Using the same downward \Ef~shift of 0.14 eV, we obtain an LDA FS topology quite comparable to the measurement at X, as shown in the lower left quadrant in Fig.~\ref{fig:fig1}(b). However, we should point out that a very small \Ef~shift of 0.01 eV upwards is needed at $\Gamma$ to match the experimental FS as shown in Fig.~\ref{fig:fig1}(b). LDA predicts 3 nearly degenerate hole bands at $\Gamma$ that cross \Ef, which cannot be resolved from our data. Assuming this 3-fold degeneracy, we find a total hole-like area of $2.7\%\times3=8.1\%$ of the BZ at $\Gamma$ and a total electron-like area of $6.9\%$ of the BZ at X. This totals to a Luttinger volume of 0.02 electrons per BZ, very close to the nominal doping level for undoped \undoped~within experimental uncertainties, compared to the much larger discrepancy of nearly 1 electron in LaOFeP~\cite{lu}.

\begin{figure*}[t]
\includegraphics[width=0.98\textwidth]{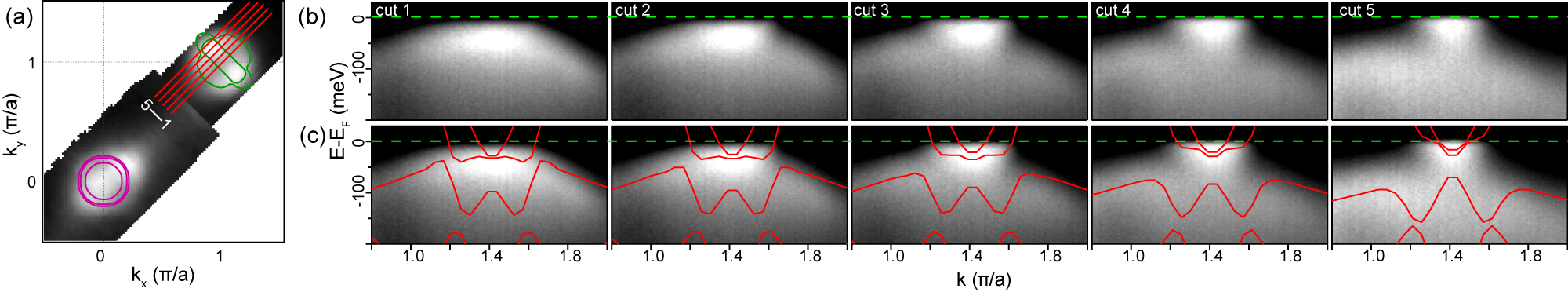}
\caption{\label{fig:fig4}ARPES spectra of \Co~(T$_c$ = 25 K) taken at 35 K ($h\nu$ = 25 eV). (a) Measured FS, overlaid with LDA FS ($k_z$=0) calculated for \undoped~after a shift of the \Ef~down by 0.13 eV at X and up by 0.03 eV at $\Gamma$. (b) ARPES spectra along five cuts marked in (a). (c) LDA bands ($k_z$ = 0) for the corresponding cuts, with \Ef~shifted down by 0.13 eV and band width renormalized by a factor of 1.4.}
\end{figure*}

The agreement between experimental data and LDA calculations is not limited to the parent compound \undoped. In Fig.~\ref{fig:fig3} and Fig.~\ref{fig:fig4}, we show similar measurements on hole doped \K~and electron doped \Co, respectively. Both compounds are optimally doped so that the SDW and structural transitions are suppressed~\cite{Kdoped,chu}. Fig.~\ref{fig:fig3}(a) shows FS map of \K~taken at 45 K in its normal state. A series of cuts parallel to the $\Gamma$-X direction (Fig.~\ref{fig:fig3}(b)) clearly shows the hole-like dispersion and the breakage into three segments due to its hybridization with the electron band away from the high symmetry line. The electron bands are again suppressed in this polarization geometry, and only appear in the perpendicular cuts (not shown) as in \undoped. Moreover, the center piece is always above the side pieces, yet always remains below \Ef. Hence, even in the hole-doped compound up to optimal doping, we confirm that only the electron bands cross \Ef, resulting in two electron pockets at X. Again, LDA calculated for \K~can match measured band dispersions in the region across X well after shifting the \Ef~down by 0.09 eV and renormalizing by a factor of 2.7, shown in Fig.~\ref{fig:fig3}(c). It is interesting to note that the renormalization effect is stronger in this compound near optimal doping. Fig.~\ref{fig:fig3}(a) shows LDA FS with a downward \Ef~shift of 0.09 eV at X and no shift at $\Gamma$, overlaid on measured FS.

Similarly, Fig.~\ref{fig:fig4}(a) shows an FS map measured on \Co~in its normal state at 35 K, and Fig.~\ref{fig:fig4}(b) shows a series of cuts parallel to the $\Gamma$-X direction. Again, we see a nearly complete hole band on the high symmetry cut, and strongly hybridized pieces off the high symmetry line. The center piece in this compound is lower than both the undoped and hole-doped compounds as we can see more of the electron pocket below \Ef, consistent with the compound being electron-doped. Since the doping for this compound is quite small, we compare its band dispersions with undoped LDA, in Fig.~\ref{fig:fig4}(c), where the \Ef~is shifted down by 0.13 eV and renormalized by a factor of 1.4. Overlaid on the measured FS in Fig.~\ref{fig:fig4}(a) is LDA FS after a momentum-dependent \Ef~shift down by 0.13 eV at X and up by 0.03 eV at $\Gamma$~\cite{co}.

Comparing across all three dopings, we see that the hole pockets at $\Gamma$ indeed becomes smaller towards electron-doping as expected. FS topology at X point is more complicated. Since many bands hybridize close to \Ef~at X, the FS topology there changes more dramatically with doping compared with $\Gamma$. But overall, the degree to which a simply shifted and renormalized LDA band structure can match well with measured band dispersion through a wide doping range in the \undoped~family is remarkable, including the number of bands, band character, and hybridization effects. One remaining mystery is the origin of the momentum-dependent shift to obtain such a match. One possibility is that the lack of charge-neutral layers in these crystals causes atoms on the cleaved surface to relax, perturbing the Fe-As bond angle that is found to be important in setting the electronic structure in these compounds~\cite{calderon}.

The results we have presented resolve a number of inconsistencies in the interpretation of ARPES results reported so far on the 122 iron pnictides. LDA predicts two electron pockets at X~\cite{singh}, but previous ARPES papers have reported intensity spots~\cite{zhouPG}, or a hole-like propeller shape at X~\cite{borisenko}. Since the hole band ($h_1$) is very flat and close to \Ef, it would contribute high intensity in the integration window when making a FS map, especially if resolution issue is considered. Furthermore, in doped samples with a T$_c$, when a superconducting gap opens at \Ef, it gaps out the spectral weight of the hybridized electron pocket that crosses \Ef, effectively increasing the relative contribution of the flat hole band, resulting in the appearance of hole-like propeller shaped FS pieces, even though this band does not cross \Ef. Hence, it is important to map out the FS above T$_c$ to avoid such complications. It has also been proposed that a folding between $\Gamma$ and X results in a gap between electron and hole bands at X~\cite{borisenko}. However, as we have shown here, the gap mentioned in that work could be alternatively interpreted as simple electron and hole band hybridization predicted by LDA, which is stronger away from the high symmetry line. It is hard to reconcile such kind of behavior with a simple band folding picture. The degree to which shifted and renormalized LDA calculation matches measured band dispersion even away from the high symmetry line strongly suggests that this is a simple band structure feature rather than more complicated many-body effects.

In summary, we have observed in the paramagnetic state 2 hole bands and 2 electron bands at X that are common features in the \undoped~family. The hole bands approach but remain below \Ef~while the electron bands cross \Ef, forming two electron pockets as the only FS sheets at X. Moreover, we found that the hole bands hybridize with one of the electron bands away from the $\Gamma$-X high symmetry line. Although more work is needed to fully understand the origin of the momentum-dependent shift and renormalization, the degree to which details in the LDA can capture the measured band dispersion and FS topology across doping regimes in ARPES data up to date suggest that this is the most appropriate and comprehensive interpretation which provides a consistent picture for the electronic structure of pnictides.

\begin{acknowledgments}
We thank I.I. Mazin, Y. Yin, H. Yao, W.S. Lee, and B. Moritz for helpful discussions. This work is supported by the U.S. Department of Energy, Office of Basic Energy Sciences at ALS (DE-AC02-05CH11231), SSRL, Stanford University (DE-AC02-76SF00515), and at ORNL. MY thanks the NSF Graduate Research Fellowship for support.
\end{acknowledgments}



\begin{references}
\bibitem{rotter} M. Rotter $et~al.$, Phys. Rev. B 78, 020503(R) (2008).
\bibitem{delacruz} C. de la Cruz $et~al.$, Nature 453, 899 (2008).
\bibitem{huang} Q. Huang $et~al.$, Phys. Rev. Lett. 101, 257003 (2008).
\bibitem{rotterdoped} M. Rotter, M. Tegel, and D. Johrendt, Phys. Rev. Lett. 101, 107006 (2008).
\bibitem{chen} H. Chen $et~al.$, Europhys. Lett. 85, 17006 (2009).
\bibitem{ni} N. Ni $et~al.$, Phys. Rev. B 78, 214515 (2008).
\bibitem{chu} J.-H. Chu $et~al.$, Phys. Rev. B 79, 014506 (2009).
\bibitem{fang} J. Dong $et~al.$, Europhys. Lett. 83, 27006 (2008).
\bibitem{singh} D.J. Singh, Phys. Rev. B 78, 094511 (2008).

\bibitem{feng} L.X. Yang $et~al.$, arXiv:0806.2627.
\bibitem{kaminski} C. Liu $et~al.$, Phys. Rev. Lett. 101, 177005 (2008). 
\bibitem{zhouspots} H. Liu $et~al.$, Phys. Rev. B 78, 184514 (2008). 
\bibitem{zhouPG} L. Zhao $et~al.$, Chin. Phys. Lett. 25, 4402-4405 (2008). 
\bibitem{ding} H. Ding $et~al.$, Europhys. Lett. 83, 47001 (2008). 
\bibitem{borisenko} V.B. Zabolotnyy $et~al.$, Nature 457, 569 (2009).
\bibitem{dingK} H. Ding $et~al.$, arXiv:0812.0534.
\bibitem{hasan} D. Hsieh $et~al.$, arXiv:0812.2289.
\bibitem{dingCo} Y. Sekiba, $et~al.$, arXiv:0812.4111.
\bibitem{lu} D.H. Lu $et~al.$, Nature 455, 81 (2008). 

\bibitem{Kdoped} G.F. Chen $et~al.$, Phys. Rev. B 78, 224512 (2008).


\bibitem{kz} Different photon energies probe the electronic structure at different $k_z$ values, with 40 eV close to $k_z$=$\pi$ while 27 eV close to $k_z$=0.
\bibitem{co} The $\Gamma$ point hole bands are very close to their band top at this doping, hence the integrated FS is more patch-like. The amount of \Ef~shift is determined from measured band dispersions.
\bibitem{calderon} M.J. Calderon, B. Valenzuela, and E. Bascones, arXiv:0810.0019
\end{references}
\end{document}